\documentclass[twocolumn]{aastex701}
\renewcommand{\added}[1]{#1}

\usepackage{amsmath} 
\usepackage{pagecolor}
\usepackage{adjustbox}
\usepackage{colortbl}   
\usepackage{array}      
\usepackage[table]{xcolor}
\usepackage{tabularx}
\usepackage{nicefrac}
\usepackage{makecell}

\definecolor{colPre}{HTML}{7AB8FF}
\definecolor{colIng}{HTML}{7FD3A7}
\definecolor{colEcl}{HTML}{F6D96B}
\definecolor{colEgr}{HTML}{F2A97F}
\definecolor{colPost}{HTML}{E39EC1}
\begin{document}

\title{Kinematically Resolving the Fe~K Complex in Her~X-1: The Accretion Disk and Ionized Wind\\Across X-ray Eclipses}

\author[orcid=0009-0007-8032-3641,sname='Sakamoto']{Koh Sakamoto}
\affiliation{Department of Physics, Graduate School of Science, Kyoto University, Kitashirakawa Oiwake-cho, Sakyo-ku, Kyoto, 606-8502, Japan}
\email[show]{sakamoto@cr.scphys.kyoto-u.ac.jp}  

\author[orcid=0000-0003-1244-3100,gname=Teruaki, sname='Enoto']{Teruaki Enoto} 
\affiliation{Department of Physics, Graduate School of Science, Kyoto University, Kitashirakawa Oiwake-cho, Sakyo-ku, Kyoto, 606-8502, Japan}
\affiliation{RIKEN Center for Advanced Photonics, 2-1 Hirosawa, Wako, Saitama 351-0198, Japan}
\email{enoto.teruaki.2w@kyoto-u.ac.jp}

\author[orcid=0000-0003-4511-8427, gname=Peter,sname=Kosec]{Peter Kosec}
\affiliation{Center for Astrophysics — Harvard \& Smithsonian, Cambridge, MA, USA}
\email{peter.kosec@cfa.harvard.edu}

\author[orcid=0000-0002-5504-4903, gname=Takeshi GO,sname=Tsuru]{Takeshi GO Tsuru}
\affiliation{Department of Physics, Graduate School of Science, Kyoto University, Kitashirakawa Oiwake-cho, Sakyo-ku, Kyoto, 606-8502, Japan}
\email{tsuru@cr.scphys.kyoto-u.ac.jp}

\author[orcid=0009-0006-7889-6144, gname=Takuto,sname=Narita]{Takuto Narita}
\affiliation{RIKEN Tamagawa High Energy Astrophysics Laboratory, 2-1 Hirosawa, Wako, Saitama 351-0198, Japan}
\email{takuto.narita@riken.jp}

\author[orcid=0009-0005-4574-7153, gname=Daijiro,sname=Hitotsuyanagi]{Daijiro Hitotsuyanagi}
\affiliation{Department of Physics, Graduate School of Science, Kyoto University, Kitashirakawa Oiwake-cho, Sakyo-ku, Kyoto, 606-8502, Japan}
\email{hitotsuyanagi.daijiro.23f@st.kyoto-u.ac.jp}

\author[orcid=0000-0003-2663-1954, gname=Laura,sname=Brenneman]{Laura Brenneman}
\affiliation{Center for Astrophysics — Harvard \& Smithsonian, Cambridge, MA, USA}
\email{lbrenneman@cfa.harvard.edu}

\author[orcid=0000-0003-1498-1543, gname=R\"udiger,sname=Staubert]{R\"udiger Staubert}
\affiliation{Institut f\"ur Astronomie und Astrophysik, Universit\"at T\"ubingen, Sand 1, D-72076 T\"ubingen, Germany}
\email{staubert@astro.uni-tuebingen.de}

\author[orcid=0000-0003-2869-7682, gname=Jon,sname=M. Miller]{Jon M. Miller}
\affiliation{Department of Astronomy, University of Michigan, Ann Arbor, MI 48109, USA}
\email{jonmm@umich.edu}

\author[orcid=0000-0002-9378-4072, gname=Andrew,sname=Fabian]{Andrew Fabian}
\affiliation{Institute of Astronomy, Madingley Road, Cambridge, CB3 0HA, UK}
\email{acf@ast.cam.ac.uk}

\author[orcid=0000-0002-5359-9497, gname=Daniele,sname=Rogantini]{Daniele Rogantini}
\affiliation{Department of Astronomy and Astrophysics, University of Chicago, 5640 S Ellis Ave, Chicago, IL 60637, USA}
\email{danieler@uchicago.edu}

\begin{abstract}
We present XRISM/Resolve spectroscopy of Her~X-1 across three X-ray eclipses observed in September 2024, \added{resolving its iron K complex through the ingress, mid-eclipse, and egress phases.}
The 5~eV high energy resolution of Resolve enabled us to disentangle and detect all primary components of the iron K complex: neutral iron fluorescence (Fe K$\alpha$ and K$\beta$), highly ionized emission lines (Fe XXV He$\alpha$ and Fe XXVI Ly$\alpha$). The neutral Fe K$\alpha$ emission is not significantly detected during mid-eclipse, indicating a compact origin near the neutron star. At ingress and egress, the line centroid exhibits red- and blue-shifts of $\sim 200$~km~s$^{-1}$ after correcting for the systemic velocity and the neutron star's orbital motion. This residual shift corresponds to Keplerian rotation at a characteristic radius of $r_{\rm disk} \sim 6.6\times10^{6}$~km, suggesting an association with the outer accretion disk.
In contrast, the highly ionized Fe~XXV~He$\alpha$ and Fe~XXVI~Ly$\alpha$ lines remain visible during eclipses, indicating an extended origin. Photoionization modeling (SPEX \texttt{pion} model) yields $\log_{10}(\xi/{\rm erg~cm~s^{-1}}) \sim 3.4$ and $N_{\rm H} \sim 3.1\times10^{22}~{\rm cm^{-2}}$ consistent with the ionized disk wind of Her~X-1. Flux-ratio diagnostics constrain the geometric inner boundary of the clumpy disk wind to $R_{\rm in} = 3^{+5}_{-2} \times 10^{10}$~cm ($1\sigma$), consistent with the Compton-heated thermal winds. The inferred mass outflow rate is $\dot{M}_{\rm out} \approx 3.2 \times 10^{-9}~M_{\odot}~{\rm yr^{-1}}$ (half the supplied mass), consistent with absorption line measurements of the disk wind obtained out of eclipse.
\end{abstract}

\keywords{\uat{Accretion}{14} --- \uat{Eclipsing binary stars}{444} --- \uat{High Energy astrophysics}{739} --- \uat{Neutron stars}{1108} --- \uat{Stellar winds}{1636} --- \uat{X-ray binary stars}{1811}}

\section{Introduction} \label{sec1.1}
Hercules~X-1 (Her~X-1) is one of the most extensively studied accreting neutron-star (NS) X-ray binaries \citep{Tananbaum1972, Giacconi1973, Reynolds1997} and has long served as a benchmark system for understanding disk accretion, X-ray reprocessing, and atmospheric or wind structures in high-inclination systems. The system comprises a magnetized neutron star with a 1.24~s spin period, orbiting its optical companion, HZ Her, every 1.700 d. At a high inclination of approximately 85$^\circ$, the system produces deep X-ray eclipses \citep{Staubert2009}. 
Her~X-1 accretes via Roche-lobe overflow, fueling a geometrically thin accretion disk around the neutron star \citep{Reynolds1997}. Table~\ref{info} summarizes the system parameters adopted in the present study.

\begin{table*}[t]
\footnotesize
\setlength{\tabcolsep}{4pt}
\renewcommand{\arraystretch}{1.15}
\centering
\caption{Summary of the system parameters of Her~X-1 and the XRISM observation.}
\begin{tabular*}{\textwidth}{l@{\dotfill}cll}
\hline\hline
\multicolumn{1}{l}{Parameter} & \multicolumn{1}{c}{Symbol} & \multicolumn{1}{l}{Value} & \multicolumn{1}{l}{Reference} \\
\hline
\multicolumn{4}{l}{\textbf{Binary system parameters}}\\
Mass of the primary star & $M_{\ast}$ & $2.40\pm0.6\,M_{\odot}$ & \citet{Leahy2025}\\
Radius of the primary star & $R_{\ast}$ & $3.92\pm 0.16\,R_{\odot} = (2.73\pm 0.11)\times10^{11}\,\mathrm{cm}$ & \citet{Leahy2025}\\
Spectral type of the primary star & -- & A/F &
\citet{Crampton1974}\\
Surface temperature of the primary star & $T_{\ast}$ & $7790 \pm 70\,\mathrm{K}$ &
\citet{Leahy2014}\\
Bolometric luminosity of the primary star & $L_{\ast}$ & $\sim5.4\times10^{1}L_{\odot}=2.1\times10^{35}\,\mathrm{erg\,s^{-1}}$ &
derived using the Stefan–Boltzmann law\\
Mass of the NS & $M_{\mathrm X}$ & $1.60\pm0.09\,M_{\odot}$ &
\citet{Leahy2025}\\
Spin period of the NS & $P_{\rm S}$ & $1.23770\,\mathrm{s}$ & This work (See also \citealt{Enoto2008}) \\
Orbital period at MJD 60563 & $P_{\rm orb}$ &
$1.70016690\pm0.00000002$ days & \citet{Staubert2009}\\
Superorbital period & $P_{\rm sup}$ & $34.8\pm 1.1 \mathrm{days}$ & \citet{Leahy2020}\\
Binary separation & $a$ & $9.52\,R_{\odot}=6.62\times10^{11}\,\mathrm{cm}$ &
derived using Kepler's third law\\
Lagrange point $L_{1}$ distance & -- & $\sim3.95\,R_{\odot}$ &
derived using \citet{Eggleton1983}\\
Inclination angle & $i$ & $(85.1\pm0.3)^\circ$ &
\citet{LeahyFrost2025}\\
Distance & $d$ & $6.6\pm0.4\,\mathrm{kpc}$ &
\citet{Reynolds1997}\\
Systemic velocity & $\gamma$ & $-65\pm2\,\mathrm{km\,s^{-1}}$ &
\citet{Reynolds1997}\\
Position (J2000) & (RA, Dec) & $(254.4576^\circ,+35.3424^\circ)$ &
\citet{Doxsey1973}\\
\hline
\multicolumn{4}{l}{\textbf{Observation summary of XRISM}} \\
ObsID & -- & 201074010 & \\
Start and end of observations & -- &
2024-09-10 02:19:06 to 2024-09-14 10:46:03& \\
Observation duration (orbital phase) & $\Phi_{\rm orb}$ &
$0.02$ to $2.57$ & \\
Total exposure time & -- &
$210\,$ks & \\
\hline
\multicolumn{4}{l}{$L_{\odot}\approx 3.83\times10^{33}~{\rm erg/s}$, $M_{\odot}\approx 1.99\times10^{30}~{\rm kg}$, $R_{\odot}\approx 6.96\times10^{10}~{\rm cm}$}\\
\end{tabular*}
\label{info}
\end{table*}

In addition to the 1.7-day orbital modulation, Her~X-1 exhibits a 35-day superorbital cycle attributed to accretion-disk precession \citep{Katz1973, GerendBoynton1976}. During the bright Main-on state, which lasts for $\sim$10 days, the spectrum shows a strong continuum and highly ionized Fe XXV He$\alpha$ and Fe XXVI Ly$\alpha$ absorption lines, produced by X-ray illumination of ionized plasma in the disk wind \citep{Reynolds1997, Leahy2001a, Kosec2020, Kosec2025}.

The light curve shows orbital-phase-dependent absorption dips. The pre-eclipse dip is caused by obscuration as the accretion stream crosses the line of sight \citep{Crosa1980, Leahy1997a, ScottLeahy1999}, while an anomalous dip at $\Phi_{\rm orb} \approx 0.4$--0.6 is linked to stream–disk interactions or outer-disk structure \citep{Igna2011}, where $\Phi_{\rm orb}$ is the orbital phase defined to be zero at mid-eclipse.

\added{X-ray eclipses span $\Delta\Phi_{\rm orb}\sim0.135$ ($\sim5.5$~hr), during which the neutron-star continuum is fully occulted \citep{Tananbaum1972, Giacconi1973}, so that the extended material reprocessing its X-rays can be observed directly. Even at mid-eclipse, roughly 1\% of the uneclipsed Main-on flux persists, attributed to scattering by extended material larger than the companion, as seen in the ASCA eclipse spectrum \citep{Choi1997}. High-resolution spectroscopy has since shown that this reprocessing gas comprises a photoionized disk atmosphere and corona, seen in recombination and Fe~K$\alpha$/K$\beta$ emission \citep{JimenezGarate2005}, together with the highly ionized disk wind detected in absorption during the bright phase \citep{Kosec2020, Kosec2025}. Such high-resolution studies have concentrated mostly on the bright phase, where the strong continuum makes absorption and emission diagnostics most effective. During eclipse, the same ionized wind that appears in absorption in the bright phase is instead seen in emission: whereas absorption constrains only the line-of-sight column and velocity, the eclipse emission reflects the full structure and thus provides a geometrically independent constraint. Indeed, in this work we find that Fe~XXV and Fe~XXVI switch from bright-phase absorption to emission during eclipse across egress (Figure~\ref{sp}; Section~\ref{sec3}). Using XRISM/Resolve, we exploit this to locate the outer-disk fluorescing region from the Doppler shift of the neutral Fe~K$\alpha$ line at ingress and egress, and to constrain the launching inner radius of the disk wind from the eclipse emission-line flux ratios.}

\section{Observation and Data Reduction} 
\label{sec2}
\begin{figure*}[ht!]
\centering
\plotone{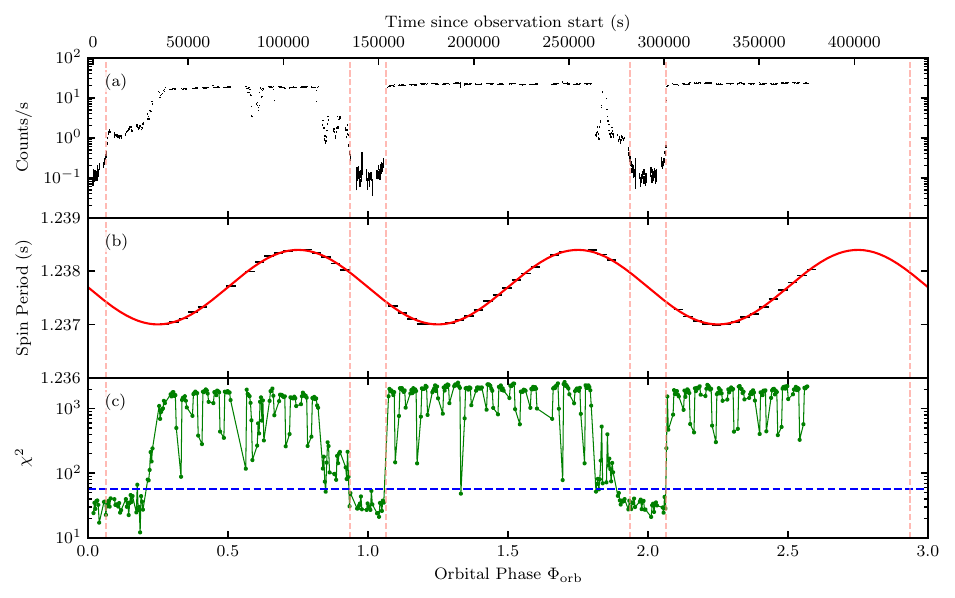}
\plotone{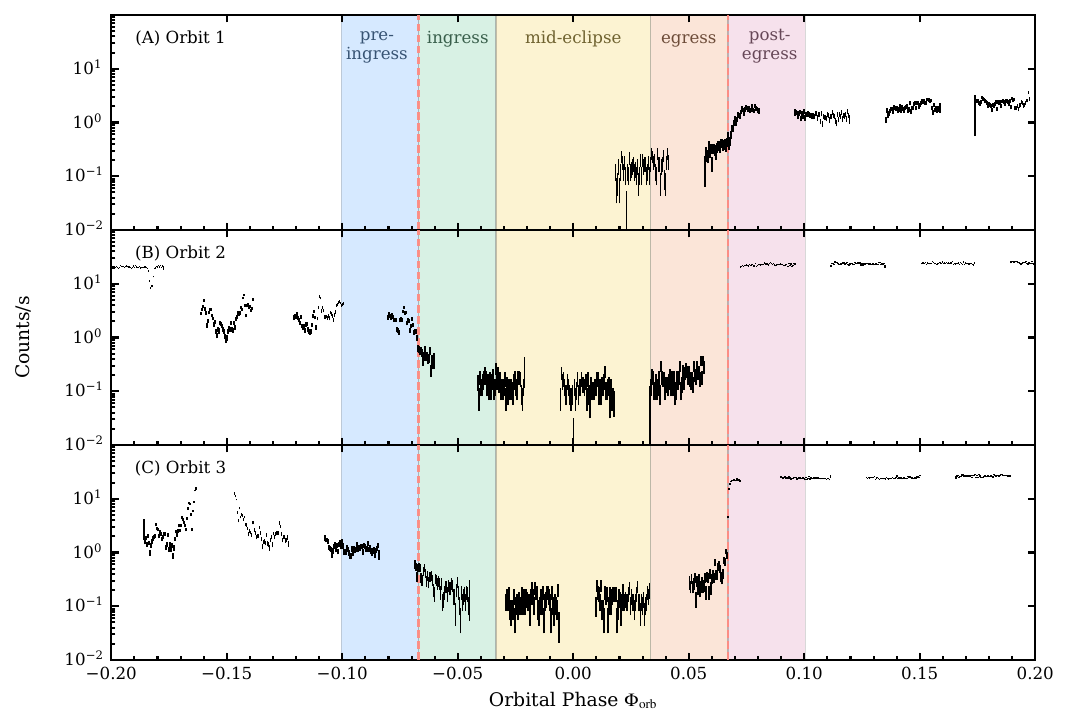}
\caption{(a) The 2.0–10 keV light curve for the XRISM Hp grade during the campaign, with data binned at 64~s intervals.
(b) Temporal evolution of the NS spin period. Intervals during eclipses and dips are excluded from this analysis. 
(c) Temporal evolution of the pulsation significance, quantified by the $\chi^2$ statistic ($\chi^2$). After folding the data at the spin periods derived in panel (b), the resulting pulse profiles are tested against the null hypothesis of a non-pulsating flat waveform. The blue dashed line indicates the 3$\sigma$ significance threshold ($\chi^2 = 57.4$), derived from a $\chi^2$ distribution with 31 degrees of freedom (corresponding to 32 phase bins). Intervals where $\chi^2$ falls below this threshold are defined as eclipse phases. For Orbit 1, however, the neutron star is observed during the low state of the 35-day superorbital cycle, suppressing pulsed emission even outside the eclipse. The eclipse boundaries are therefore defined by the orbital phases derived from Orbits 2 and 3.
Panels (A)--(C) display zoomed light curves around the three eclipses, labeled Orbit 1, Orbit 2, and Orbit 3 in chronological order. The horizontal axis represents orbital phase $\Phi_{\rm orb}$, with mid-eclipse defined as phase 0. The red vertical dashed lines indicate ingress and egress, and each eclipse interval is divided into five color-coded segments: pre-ingress, ingress, mid-eclipse, egress, and post-egress, with relative durations of 1:1:2:1:1.}
\label{lc}
\end{figure*}

XRISM observed Her~X-1 starting on 2024 September 10 (ObsID 201074010, GO1), with a total on-target duration of 380 ks and a net Resolve exposure of 210 ks \citep{Kosec2025}. Throughout the observation, the gate valve remained closed, which attenuated the low-energy effective area below $\sim$2 keV while leaving the Fe K band fully accessible. 
\autoref{lc}a displays the 2–10 keV light curve extracted from the Resolve event data.
The XRISM observation covers approximately 2.5 orbital cycles, commencing midway through the eclipse of the first binary orbit.

We focus on the eclipse and its adjacent phases, where the occultation of the bright direct continuum provides a low-background environment ideal for identifying faint emission lines.
Analysis of the out-of-eclipse bright phase is presented by \citet{Kosec2025}, focusing on high-velocity disk winds, while the broadened Fe~K emission line was investigated by \citet{Kosec2026}, and a detailed pulse-phase-resolved spectroscopy is planned to be provided by T.~Narita et al. (in prep.).

The XRISM data were reduced using HEASoft v6.35.1 \added{\citep{Heasoft2014}} in conjunction with the CALDB version 20250325, following the standard procedures described in the XRISM ABC Data Reduction Guide\footnote{https://heasarc.gsfc.nasa.gov/docs/xrism/analysis/}. Events were extracted and used from the full Resolve array, excluding pixels 12 (the calibration pixel) and 27 (due to known gain instabilities).
The analysis was restricted to high-resolution primary (Hp) events.
Event times were corrected to the solar-system barycenter using \texttt{barycorr}, with the spacecraft orbit file (.orb) and the J2000 source coordinates of Her~X-1, (R.A., DEC) = (254.457546$^\circ$, +35.342357$^\circ$).

The instrumental response files were generated using \texttt{rslmkrmf} with the x-large size matrix for the response matrix files (rmf) and \texttt{xaarfgen} for the ancillary response files (arf). No background spectrum was subtracted. The standard XRISM/Resolve background model predicts a photon flux density in the 6.0–7.5 keV band that is $\sim$2\% of the Her~X-1 continuum level during eclipse, thus not affecting the present analysis. No additional spectral binning was applied. Unless otherwise stated, all errors in this paper are reported 
at the 1$\sigma$ level.

\section{Analysis and Results}\label{sec3}
%\subsection{Orbital Phase and Eclipse Selection}\label{sec3.1}
\subsection{Orbital Ephemeris and Eclipse Segmentation}\label{sec3.1}
We first performed a pulse-timing analysis of Her~X-1 to assign orbital phases $\Phi_{\rm orb}$. The event data were divided into 5~ks intervals, and the pulse period for each interval was determined by epoch-folding of the event data, as illustrated by the periodograms in \autoref{lc}b.
The observed pulse periods show sinusoidal modulation due to the Doppler effect induced by the NS's nearly circular orbital motion.
This modulation was fitted by $P(t)=P_0 + A \sin(\omega t + \phi_0)$, where $P_0$ is the intrinsic pulse period, $A$ is the amplitude of the Doppler modulation, $\omega$ is the orbital angular frequency, and $\phi_0$ is the initial phase offset.
The best-fit parameters were $P_0=1.237700(3)$~s, $A=6.99(3) \times 10^{-4}$~s, $\omega=4.271(5) \times 10^{-5}$~rad/s, and $\phi_0=3.25(1)$~rad. This model (red solid line in \autoref{lc}b) allows for precise determination of $\Phi_{\rm orb}$, where we define the superior conjunction (mid-eclipse) as phase $\Phi_{\rm orb}=N$ (where $N$ is an integer) with $\Phi_{\rm orb}=0$ corresponding to MJD(TDB)$=$60563.066(2).

To define the eclipse boundaries, we statistically evaluated the significance of the pulsation signal. The data were grouped into 500~s intervals and the $\Delta \chi^2$-statistic was calculated as the deviation from a flat (non-pulsing) null hypothesis. As shown in \autoref{lc}c, the eclipse was determined by the disappearance of the pulse signal, defined as the time span where $\Delta \chi^2$ drops below the $3\sigma$ significance threshold ($\Delta \chi^2 = 57.4$). This transition is consistent with the flux decrement shown in \autoref{lc}a. 
Based on this analysis, the full eclipse interval was defined as $\Phi_{\rm orb} \in [N-0.067, N+0.067]$. 
For phase-resolved spectroscopy, we subdivided the data into five orbital intervals with a duration ratio of 1:1:2:1:1. These intervals are defined as follows (see also \autoref{lc}A--C):
\begin{itemize}
    \item Pre-ingress: $[N-0.1005, N-0.067]$
    \item Ingress: $[N-0.067, N-0.0335]$
    \item Mid-eclipse: $[N-0.0335, N+0.0335]$
    \item Egress: $[N+0.0335, N+0.067]$
    \item Post-egress: $[N+0.067, N+0.1005]$
\end{itemize}

\subsection{Spectral Extraction and Modeling Strategy}\label{sec3.2}
\begin{figure*}[ht!]
\centering
\plotone{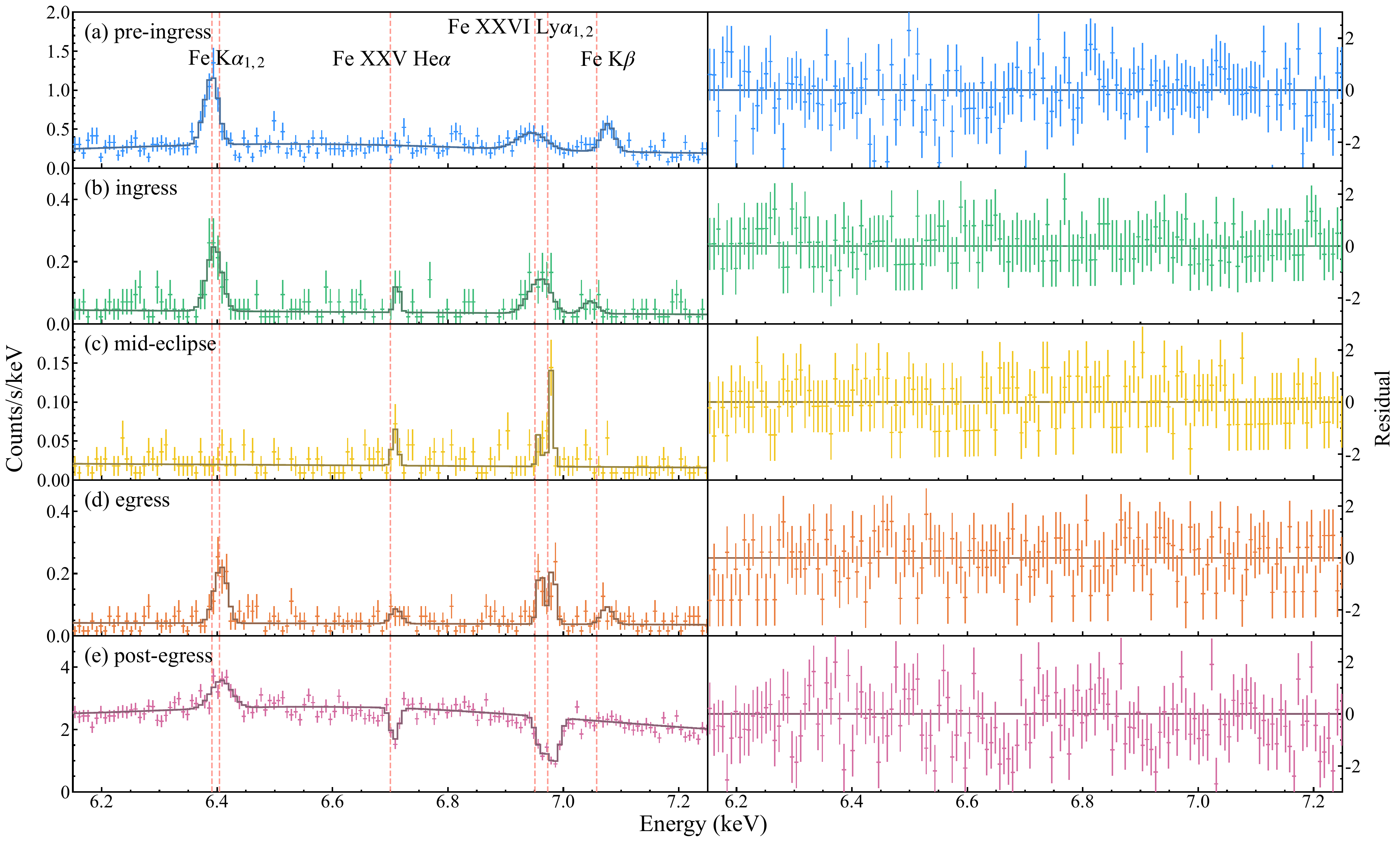}
\caption{XRISM/Resolve phase-resolved spectra of Her~X-1 in the 6.2–7.2 keV range for five orbital-phase intervals: (a) pre-ingress, (b) ingress, (c) mid-eclipse, (d) egress, and (e) post-egress. For each phase interval, spectra from the three eclipses were co-added.
Colored points represent the observed spectra and the solid lines indicate the best-fit models.
Vertical dashed lines indicate the rest-frame energies of Fe K$\alpha$ (6.404, 6.391~keV), Fe XXV He$\alpha$ (6.700~keV), Fe XXVI Ly$\alpha$ (6.952, 6.973~keV), and Fe K$\beta$ (7.058~keV) lines. The corresponding residuals (data/model ratio) for each fit are shown in the right-hand sub-panels.}
\label{sp}
\end{figure*}

\autoref{sp} displays the phase-resolved XRISM/Resolve spectra in the Fe--K band, which were extracted for the five orbital intervals defined in Section~\ref{sec3.1} by co-adding data from the three observed orbital cycles.
The high energy resolution of Resolve clearly disentangles the Fe K complex into four primary components: neutral Fe K$\alpha$ and K$\beta$ fluorescence emissions, highly ionized Fe XXV He$\alpha$ lines, and Fe XXVI Ly$\alpha$ lines. \added{With the gate valve closed, the effective area below $\sim$2~keV is strongly attenuated (Section~\ref{sec2}), and within the accessible band no emission lines from elements other than iron are significantly detected during the eclipse; we therefore focus our analysis on the Fe~K complex.}

Spectral modeling was conducted in two steps. First, we established a stable baseline by fitting the 3--10~keV continuum with a power-law model, excluding the 6.0--7.5~keV Fe-complex region. This continuum was then frozen to model the Fe K emission and absorption features within the 6.0--7.5~keV band. The specific model components were adjusted for each interval to best represent the observed features, and the resulting best-fit parameters are summarized in \autoref{par}.

For the ingress, mid-eclipse, and egress intervals (\autoref{sp}b, c, d), four narrow Fe-line components were modeled: Fe K$\alpha$, Fe K$\beta$, Fe XXVI Ly$\alpha$, and Fe XXV He$\alpha$ complex. Each line component was represented by one or two Gaussian functions, depending on the intrinsic multiplet structure. The centroids of the multiplet Gaussians of the same ionic species were coupled via a common Doppler shift (redshift parameter), which was allowed to vary independently across each interval.

For the pre-ingress interval (\autoref{sp}a), the model included two additional components: a broad Fe K emission and an Fe K-edge. The broad emission, modeled as a redshifted broadened Gaussian, is attributed to reprocessing in the inner disk or accretion column near the neutron star and disappears during the deeper eclipse phases (P.~Kosec et al., in prep.; T.~Narita et al., in prep.). In these cases, the broad Fe K feature and the Fe K-edge were constrained simultaneously with the continuum in the 3--10~keV band and subsequently fixed before modeling the narrow Fe features in the 6.0--7.5~keV range. The Fe K-edge is attributed to excess photoelectric absorption associated with the pre-eclipse dip, which preferentially occurs before eclipses and is absent in the post-egress interval.

In the post-egress interval (\autoref{sp}e), the highly ionized Fe XXV He$\alpha$ and Fe XXVI Ly$\alpha$ features appear in absorption rather than in emission. These were modeled with multiplicative Gaussian absorption profiles (\texttt{gabs} in XSPEC), with their multiplet components tied through a common redshift parameter for each ion. Similar to the pre-ingress case, the continuum and broad Fe K emission were fixed after being constrained in the 3--10~keV band; notably, the Fe K-edge was not required for the post-egress spectrum.

\begin{table*}[ht!]
{\centering
\scriptsize
\caption{Best-fit parameters of the Fe-line features across the five phase intervals in \autoref{sp}.}
\label{par}
\begin{tabular}{lll|lllll}
\hline
&&& {\bfseries\color{colPre!45!black}(a) pre-ingress} & {\bfseries\color{colIng!45!black}(b) ingress} & {\bfseries\color{colEcl!45!black}(c) mid-eclipse} & {\bfseries\color{colEgr!45!black}(d) egress} & {\bfseries\color{colPost!45!black}(e) post-egress}\\
components & parameters & units & {\bfseries\color{colPre!45!black}\scalebox{0.5}{\large$-0.1005\leq\Phi_{\rm orb}<-0.067$}} & {\bfseries\color{colIng!45!black}\scalebox{0.5}{\large$-0.067\leq\Phi_{\rm orb}<-0.0335$}} & {\bfseries\color{colEcl!45!black}\scalebox{0.5}{\large$-0.0335\leq\Phi_{\rm orb}<0.0335$}} & {\bfseries\color{colEgr!45!black}\scalebox{0.5}{\large$0.0335\leq\Phi_{\rm orb}<0.067$}} & {\bfseries\color{colPost!45!black}\scalebox{0.5}{\large$0.067\leq\Phi_{\rm orb}\leq0.1005$}} \\
\hline
$\rm powerlaw$  & PhoIndex   & & $-0.31^{+0.05}_{-0.05}$ & $0.7^{+0.1}_{-0.1}$ & $0.79^{+0.09}_{-0.09}$ & $0.51^{+0.08}_{-0.08}$ & $0.87^{+0.01}_{-0.01}$ \\
                & Norm$^{\dag}$ & & $8.8^{+0.9}_{-0.8} \times 10^{-4}$ & $1.1^{+0.2}_{-0.2} \times 10^{-3}$ & $6.3^{+1.0}_{-0.9} \times 10^{-4}$ & $8^{+1}_{-1} \times 10^{-4}$ & $9.1^{+0.2}_{-0.2} \times 10^{-2}$ \\
$\rm Fe K\alpha$ & LineE      & keV & 6.404, 6.391(fixed) & 6.404, 6.391(fixed) & 6.404, 6.391(fixed)$^{\ddag}$ & 6.404, 6.391(fixed) & 6.404, 6.391(fixed) \\
                & Redshift$^{\ast}$   & km/s& $450_{-70}^{+70}$ & $210_{-130}^{+130}$ & $-69$(fixed)$^{\ddag}$ & $-300_{-110}^{+110}$ & $-340_{-150}^{+170}$\\
                & Sigma      & km/s& $470_{-70}^{+80}$ & $520_{-90}^{+110}$ & 470(fixed)$^{\ddag}$ & $420_{-120}^{+140}$ & $780_{-160}^{+190}$\\
                & Norm$_{1,2}^{\dag}$ & & $2.3_{-0.3}^{+0.3} \times 10^{-4}$ & $5.7_{-1.0}^{+1.1} \times 10^{-5}$ & $\leq4.4\times 10^{-6}{}^{\ddag}$ & $4.3_{-0.8}^{+0.8} \times 10^{-5}$ & $3.4_{-0.5}^{+0.5} \times 10^{-4}$\\
                & EW         & eV  & 91 & 164 & $\leq25^{\ddag}$ & 128 & 15\\
$\rm Fe K\beta$  & LineE      & keV & 7.058(fixed) & 7.058(fixed) & ––– & 7.058(fixed) & ---\\
                & Redshift$^{\ast}$   & km/s& $-790_{-130}^{+140}$ & $480_{-460}^{+310}$ & ––– & $-740_{-320}^{+310}$ & ---\\
                & Sigma      & km/s& $470$(fixed) & $520$(fixed) & ––– & $420$(fixed) & ---\\
                & Norm$^{\dag}$       & & $8.8_{-1.8}^{+1.9} \times 10^{-5}$ & $1.1_{-0.6}^{+0.7} \times 10^{-5}$ & ––– & $1.2_{-0.5}^{+0.6} \times 10^{-5}$ & ---\\
                & EW         & eV  & 47 & 30 & ––– & 32 & ---\\
$\rm Fe Ly\alpha$& LineE      & keV & 6.952, 6.973(fixed) & 6.952, 6.973(fixed) & 6.952, 6.973(fixed) & 6.952, 6.973(fixed) & 6.952, 6.973(fixed) \\
                & Redshift$^{\ast}$   & km/s& $1000_{-1000}^{+500}$ & $300_{-700}^{+500}$ & $-250_{-40}^{+40}$ & $-350_{-50}^{+50}$ & $-450_{-50}^{+50}$\\
                & Sigma      & km/s& $1000_{-300}^{+1000}$ & $600_{-200}^{+300}$ & $110_{-30}^{+50}$ & $200_{-40}^{+50}$ & $350_{-40}^{+60}$\\
                & Norm$_2^{\dag}$   & & $3_{-3}^{+9} \times 10^{-5}$ & $1.6_{-1.6}^{+3.4} \times 10^{-5}$ & $3.7_{-1.6}^{+2.0} \times 10^{-6}$ & $2.1_{-0.5}^{+0.6} \times 10^{-5}$ & \\
                & (Strength$_2$) & (keV) &&&&& $1.3_{-0.2}^{+0.2} \times 10^{-2}$\\
                & Norm$_1^{\dag}$   & & $9_{-9}^{+5} \times 10^{-5}$ & $2.8_{-1.3}^{+2.2} \times 10^{-5}$ & $1.1_{-0.3}^{+0.3} \times 10^{-5}$ & $2.2_{-0.5}^{+0.6} \times 10^{-5}$ & \\
                & (Strength$_1$) & (keV) &&&&& $2.0_{-0.2}^{+0.2} \times 10^{-2}$\\
                & EW         & eV  & 56 & 149 & 97 & 133 & 33\\
$\rm Fe He\alpha$& LineE      & keV & --- &  $6.700$(fixed) & $6.700$(fixed) & $6.700$(fixed) & $6.700$(fixed) \\
                & Redshift$^{\ast}$   & km/s& --- & $-530_{-130}^{+120}$ & $-360_{-110}^{+110}$ & $-460_{-260}^{+350}$ & $-330_{-60}^{+60}$\\
                & Sigma      & km/s& --- & $170_{-170}^{+100}$ & $200_{-80}^{+100}$ & $400_{-140}^{+380}$ & $270_{-50}^{+50}$\\
                & Norm$^{\dag}$       & & --- & $1.0_{-0.5}^{+0.6} \times 10^{-5}$ & $5.2_{-2.0}^{+2.4} \times 10^{-6}$ & $9.4_{-4.3}^{+5.5} \times 10^{-6}$ & \\
                & (Strength) & (keV) &&&&& $8.2_{-1.3}^{+1.4} \times 10^{-3}$\\
                & EW         & eV  & --- & 22 & 30 & 20 & 8\\
\hline
\end{tabular}\par}

\smallskip
\small
\noindent Emission lines are modeled using Gaussian profiles (\texttt{gauss}), whereas the highly ionized Fe~\textsc{xxvi} Ly$\alpha$ and Fe~\textsc{xxv} He$\alpha$ features in the post-egress interval are modeled using multiplicative Gaussian absorption profiles (\texttt{gabs}); the tabulated values represent the absorption strength (integrated optical depth). Energies marked as ''(fixed)
'' were held constant during fitting.

\smallskip
\noindent $^{\ast}$Positive redshift indicates motion away from the observer.

\smallskip
\noindent $^{\dag}$The normalization parameter (Norm) is given in units of $\rm photons~cm^{-2}~s^{-1}$.

\smallskip
\noindent $^{\ddag}$For mid-eclipse, where Fe K$\alpha$ is not significantly detected, an upper limit is derived by including a Gaussian component with fixed centroid energy (rest-frame emission energy), fixed redshift corresponding to the systemic velocity of Her~X-1, and fixed line width (averaged from ingress and egress). 
The $1\sigma$ uncertainty of the normalization is adopted as the upper limit, and the corresponding EW upper limit is also reported.
\end{table*}

\subsection{Neutral Fe K$\alpha$ emission}\label{sec3.3}
In all intervals, we modeled the neutral Fe K$\alpha$ doublets using laboratory rest energies of $\rm E_{K\alpha1}=6.404~keV$ and $\rm E_{K\alpha2}=6.391~keV$ as the model baseline, with their observed centroid deviations attributed to the Doppler shifts discussed below.
The neutral Fe K$\beta$ line is also detected with an intensity ratio relative to the Fe K$\alpha$ line of $\sim$0.14 as expected.
The equivalent width (EW) of the Fe K$\alpha$ line exhibits a clear phase dependence, increasing from $\sim$90~eV in pre-ingress to $\sim$160~eV during ingress when the direct continuum is strongly suppressed.
During mid-eclipse the line is not detected, and the 1$\sigma$ upper limit on the equivalent width is 25 eV, well below the values measured in the adjacent phases. The line reappears at egress with an EW of $\sim$130~eV before decreasing to $\sim$15~eV in the post-egress phase, as the continuum emission recovers.

In addition to the flux modulation, the centroid energy of the Fe K$\alpha$ line exhibits a systematic phase dependence.
The inferred line-of-sight velocities are $\rm +460 \pm 70~km~s^{-1} (-9.7\pm1.5~eV)$ in pre-ingress, $\rm +210 \pm 130~km~s^{-1} (-4.5\pm2.8~eV)$ in ingress, $\rm-300 \pm 110~km~s^{-1} (+6.4\pm2.3~eV)$ in egress, and $\rm -340 \pm 150~km~s^{-1} (+7.3\pm3.2~eV)$ in post-egress.
Here, positive velocities correspond to redshift (recession from the observer).
The magnitude of these velocities exceeds the line-of-sight component of the NS orbital velocity at ingress/egress ($\Phi_{\rm orb} = \pm 0.067$), $170~{\rm km~s^{-1}} \times \sin(2\pi\times 0.067) \approx 69~{\rm km~s^{-1}}$, with $\rm 170~km~s^{-1}$ being the NS orbital velocity \citep{Deeter1981}. This indicates that the observed Doppler shifts are not dominated by the binary orbital motion. The center energy of the Fe K$\beta$ emission line has large statistical uncertainties during ingress and egress phases, making it unsuitable for determining the ionization state as discussed by \citet{Nagai2026}.

\subsection{Highly ionized Fe lines} \label{sec3.4}

In contrast to the neutral Fe fluorescence, \added{the highly ionized Fe~XXVI~Ly$\alpha$ line is detected in all orbital-phase intervals, and Fe~XXV~He$\alpha$ is marginally detected in every interval except pre-ingress}. In the post-egress interval, the highly ionized Fe features are observed as absorption lines rather than emission. Similarly, outside eclipse, they are detected as absorption lines during bright phases \citep{Kosec2025}.

In pre-ingress and ingress, the Fe XXV and Fe XXVI centroids show redshifts of $\sim 300\text{--}1000~{\rm km~s^{-1}}$ with $1\sigma$ uncertainties of $500\text{--}1000~{\rm km~s^{-1}}$, whereas in mid-eclipse, egress, and post-egress, they show blueshifts of $300\text{--}500~{\rm km~s^{-1}}$ with $1\sigma$ uncertainties of $\sim 50~{\rm km~s^{-1}}$ relative to their respective rest energies. \added{We identify the Fe~XXV~He$\alpha$ feature with the resonance line $w$ (6.700~keV): its velocity matches that of Fe~XXVI~Ly$\alpha$ to within ${\sim}100~{\rm km~s^{-1}}$, whereas the intercombination ($x, y$) or forbidden ($z$) components would be blueshifted by ${\sim}1000$--$2800~{\rm km~s^{-1}}$. This is corroborated by the bright-phase wind, where the same Fe~XXV absorption, weaker than Fe~XXVI but significantly detected, is supporting the identification with the resonance line $w$ \citep[][Table~1]{Kosec2025}.}

The detection of Fe~XXV and Fe~XXVI emission lines with velocity shifts of $\sim100\text{--}1000~{\rm km~s^{-1}}$ in the eclipse spectra suggests that we may be observing photoionized re-emission from the disk wind plasma that appears in absorption during the bright out-of-eclipse phases \citep{Kosec2025}. \added{This plasma is in photoionization equilibrium, as established for the disk atmosphere and corona by \citet{JimenezGarate2005} and confirmed for the disk wind by \citet{Kosec2020}.} To derive the physical properties of this emitting plasma and compare them with the wind observed out of eclipse, we accordingly performed self-consistent photoionization modeling in the 6.2--7.2~keV band using SPEX on the eclipse spectra (combining the ingress, mid-eclipse, and egress).
The total spectral model is
\begin{equation}
S(E) = \texttt{hot} \times \left[ \texttt{powerlaw} + \texttt{gaussian} + \texttt{pion}(\mathrm{SED}) \right],
\end{equation}
where \texttt{hot} accounts for line-of-sight absorption, \texttt{powerlaw} describes the continuum, \texttt{gaussian} models the neutral or low-ionization Fe~K$\alpha$ fluorescence line, and \texttt{pion} reproduces the highly ionized Fe~XXV and Fe~XXVI emission lines by computing the emission spectrum of a photoionized plasma in equilibrium with the local radiation field. The irradiating SED for \texttt{pion} was adopted from the broadband bright-phase model \citep{Kosec2025}, with its direct emission suppressed by \texttt{etau} to simulate the occultation of the central source during eclipse. In the \texttt{pion} component, the hydrogen column density $N_{\rm H}$, ionization parameter $\xi$, velocity width $v$, and line-of-sight velocity $z_v$ were treated as free parameters, while the covering fraction $\Omega/4\pi$ was fixed at 0.1 \citep{Kosec2020}.

The best-fit parameters obtained from the SPEX analysis indicate a highly ionized plasma with an ionization parameter of $\log (\xi/{\rm erg~cm~s^{-1}}) = 3.42^{+0.08}_{-0.07}$ and a column density of $N_{\rm H} = 3.1^{+0.9}_{-0.6} \times 10^{22}~{\rm cm^{-2}}$. \added{These values are of the same order as the wind properties independently derived from the absorption features during the bright phase \citep{Kosec2025}}, suggesting that the emission lines observed during the eclipse originate from the same clumpy disk wind that causes line-of-sight absorption outside the eclipse.

\section{Discussion} \label{sec4}

\subsection{Origin of the neutral Fe K$\alpha$ emission}\label{sec4.1}
The eclipse-phase dependence of the neutral Fe~K$\alpha$ line provides direct constraints on the spatial distribution of the fluorescing material. The non-detection of the line during mid-eclipse (Section~\ref{sec3.3}) indicates that the emitting region is fully contained within the companion star's shadow, demonstrating that the fluorescent emission is confined to within $R_{\ast}/2 \sim 1.4 \times 10^{11}~{\rm cm}$ of the neutron star. This behavior disfavors an extended or vertically distributed origin, such as a large-scale wind or the stellar atmosphere of the companion, and instead strongly supports an association with the accretion disk.

Examining the line flux across eclipse phases further constrains the spatial distribution of this emission. The Fe~K$\alpha$ normalization at ingress and egress is ${\sim}5 \times 10^{-5}~{\rm photons~cm^{-2}~s^{-1}}$, approximately 20\% of the post-egress value of ${\sim}3 \times 10^{-4}~{\rm photons~cm^{-2}~s^{-1}}$ (Table~2). For a geometrically flat disk, where the fluorescent emissivity scales as $r^{-3}$~\citep{Willkins2012}, any outer component would be negligible. The presence of a non-negligible outer contribution that persists at ingress and egress is therefore difficult to reconcile with a flat disk, and may instead point to a non-flat disk geometry, such as a flared or warped disk \citep{Katz1973}. A quantitative characterization of this outer emission requires a three-dimensional emissivity model accounting for the disk geometry, which we defer to future work.

In addition, the centroid energy of the Fe K$\alpha$ line exhibits systematic Doppler shifts across the eclipse, shifting from red in ingress to blue in egress. The magnitude of these shifts exceeds both the systemic velocity of Her~X-1 ($v_{\rm sys} = -65 \pm 2~\mathrm{km~s^{-1}}$; \citealt{Reynolds1997}) and the projected line-of-sight contribution from the neutron star orbital motion near eclipse ($v_{\rm orb, los} = 69 \pm 4~\mathrm{km~s^{-1}}$; derived from \citealt{Deeter1981}). This clearly demonstrates that the observed Doppler shifts are not dominated by the binary motion, but instead arise from the intrinsic bulk motion of the fluorescing material itself.

Assuming that these Doppler shifts originate from the rotational motion of the accretion disk, the disk rotational velocity ($v_{\rm rot}$) can be isolated by removing the contributions from the systemic velocity and the neutron star orbital motion. During ingress, the observed velocity satisfies
\begin{equation}
v_{\rm obs} = v_{\rm sys} + v_{\rm orb, los}\sin i + v_{\rm rot}\sin i,
\end{equation}
while during egress it follows
\begin{equation}
v_{\rm obs} = v_{\rm sys} - v_{\rm orb, los}\sin i - v_{\rm rot}\sin i,
\end{equation}
where $i$ is the inclination angle of the orbital plane with respect to the line of sight. Since $i \sim 85^\circ$ \citep{LeahyFrost2025}, we approximate $\sin i \approx 1$.
Based on the observed Doppler shifts, we derive disk rotational velocities of $v_{\rm rot} = 208 \pm 130~\mathrm{km~s^{-1}}$ during ingress and $v_{\rm rot} = 165 \pm 108~\mathrm{km~s^{-1}}$ during egress. These two estimates are mutually consistent within their uncertainties. Thus, we adopt a representative disk rotational velocity of $v_{\rm rot} \simeq 180~\mathrm{km~s^{-1}}$ for the following analysis.
To map this velocity to a characteristic emission radius on the disk, we assume Keplerian rotation around a neutron star with mass $M_{\rm X} = 1.6\,M_\odot$ \citep{Leahy2025}.
The representative disk rotational velocity then corresponds to
\begin{equation}
r_{\rm disk} = \frac{G M_{\rm X}}{v_{\rm rot}^2} \sim 6.6\times10^{6}~\mathrm{km}.
\end{equation}
\added{Here $r_{\rm disk}$ is an effective Keplerian radius, inferred under the assumption that the measured centroid shift traces the projected rotational velocity of the disk material.}
The outer radius of the accretion disk has been estimated to be $\sim 1.7\times10^{6}$~km from the orbital modulation of X-ray-reprocessed optical emission \citep{Howarth1983}. Our derived $r_{\rm disk}$ exceeds this by a factor of $\sim$4, which we attribute to the luminosity-weighted averaging over the extended emitting region inherent in the single-radius approximation. At the order of magnitude level, both values are consistent with emission from the outer part of the accretion disk.

The observed width of the neutral Fe K$\alpha$ line ($400\text{--}700~\rm km~s^{-1}$) reflects a combination of geometric broadening, turbulent dispersion, and bulk outflow motions that are strongly degenerate with the present data. We therefore focus on the line centroid shifts as a more direct probe of the disk bulk motion.

\begin{figure*}[ht!]
\centering
\includegraphics[width=\textwidth]{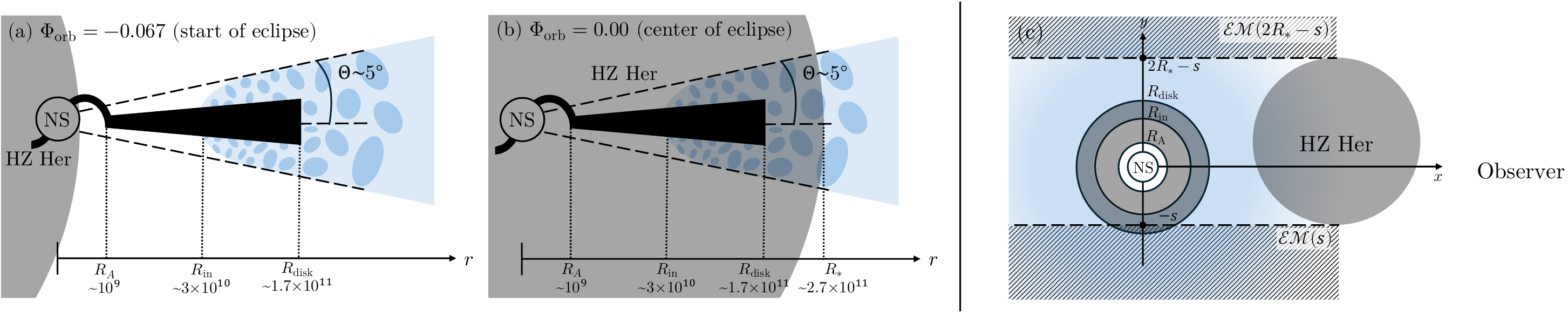}
\caption{Schematic of the Her~X-1 system geometry (not to scale): (a) side-on view at start of eclipse ($\Phi_{\rm orb}=-0.067$) and (b) center of eclipse ($\Phi_{\rm orb}=0.00$). The clumpy disk wind (blue) is launched with a biconical half-opening angle of $\sim 5^\circ$. Key radii indicated are the Alfv\'{e}n radius $R_A \sim 10^{9}$~cm, the wind inner boundary $R_{\rm in} \sim 3 \times 10^{10}$~cm, the outer edge of the accretion disk $R_{\rm disk} \sim 1.7 \times 10^{11}$~cm \citep{Howarth1983}, and the companion star radius $R_\ast \sim 2.7 \times 10^{11}$~cm. (c) Top-down view of the orbital plane, with the observer along the $x$-axis. The hatched regions indicate the portions of the wind visible to the observer during eclipse, and the observed emission line flux corresponds to $\mathcal{EM}(s)$ and $\mathcal{EM}(2R_\ast - s)$ from the near and far limbs of HZ~Her, respectively, where $s \geq 0$ is the projected distance from the neutron star to the near limb of HZ~Her.}
\label{ponti}
\end{figure*}

\subsection{Physical Properties of the Clumpy Disk Wind}\label{42}

In addition to the neutral Fe fluorescence emission, we also detected highly ionized Fe emission lines. The plasma emitting these lines has properties consistent with the ionized disk wind of Her~X-1, seen in absorption during the high-flux out-of-eclipse intervals \citep{Kosec2025}.
To understand the driving mechanism of the clumpy disk wind, we constrain its geometric inner boundary, $R_{in}$. We model the disk wind as an $r^{-2}$ density profile confined within a biconical equatorial wedge of half-opening angle $\Theta_0 \sim 5^{\circ}$ (above and below the disk; see \autoref{ponti}a,b). This profile follows naturally from the simplest assumption of a steady, constant-velocity outflow, and is consistent with the approximately constant ionization parameter observed across eclipse phases in this work, with height above the disk \citep{Kosec2023}, and between the eclipse and bright-phase intervals \citep{Kosec2025}. The detailed geometric calculation of how the emission line intensity varies as the companion star HZ~Herculis progressively occults the system is provided in \autoref{app:A}. Briefly, as HZ~Herculis moves across the line of sight, we define a masking depth $s \geq 0$ as the projected distance from the neutron star to the near limb of the companion (see \autoref{ponti}c). In this framework, the companion star masks a specific band of the wind emission region, leaving the near side and far side of the wind simultaneously visible. By comparing the observed emission line flux at each $s$ with this geometric model, we can evaluate the extent of the emission region. As derived in \autoref{app:A}, denoting the phase-averaged emission line flux as $\langle F \rangle$, this yields the analytical ratio \begin{equation}   
\eta_{\rm line} \equiv \frac{\langle F \rangle_{\rm ing+egr}}{\langle F \rangle_{\rm mid}}   = \frac{\langle \mathcal{EM} \rangle_{\rm ing}}{\langle \mathcal{EM} \rangle_{\rm mid}}   = \frac{2 + \ln\!\left(\dfrac{R_\ast}{3\,R_{in}}\right)}{\ln 3}, 
\end{equation} 
which inverts to 
\begin{equation}   
R_{in} = \frac{R_\ast}{3}\,\exp\!\left[2 - \eta_{\rm line}\cdot\ln 3\right]. 
\end{equation} 
Using the combined He$\alpha$ and Ly$\alpha$ emission line normalizations, we calculate the flux ratio of the combined ingress and egress phases to the mid-eclipse phase, $\eta_{\rm line} \equiv \langle F \rangle_{\rm ing+egr} / \langle F \rangle_{\rm mid} = 2.76_{-0.85}^{+1.19}$. This ratio translates to an inner boundary of $R_{in} = 3\,^{+5}_{-2} \times 10^{10}~{\rm cm}$ ($1\sigma$ confidence level).  

This constraint restricts the wind base to within the outer edge of the accretion disk. According to Compton-heating theory \citep{Begelman1983}, a thermal wind is launched at radii in the range $0.1\,R_{\rm IC} \lesssim R \lesssim R_{\rm IC}$, where $R_{\rm IC} = GM_X\mu m_p / kT_{\rm IC}$ is the inverse Compton radius, with $\mu = 1.4$ the mean molecular weight per hydrogen nucleus, $m_p$ the proton mass, $k$ the Boltzmann constant, and $T_{\rm IC}$ the inverse Compton temperature. For Her~X-1, assuming $kT_{\rm IC} \sim 3$~keV \citep{Kosec2020}, this corresponds to $R_{\rm IC} \sim 8 \times 10^{10}$~cm and a launching window of ${\sim}8 \times 10^{9}$--$8 \times 10^{10}$~cm. Our derived $R_{in} = 3\,^{+5}_{-2} \times 10^{10}$~cm falls well within this range, supporting a thermally driven origin for the clumpy disk wind. A schematic summary of the system geometry, including these observationally constrained characteristic radii, is illustrated in \autoref{ponti}a,b. 

\added{Furthermore, our eclipse values ($\log\xi = 3.42^{+0.08}_{-0.07}$, $N_{\rm H} = 3.1^{+0.9}_{-0.6}\times10^{22}~{\rm cm^{-2}}$; Section~\ref{sec3.4}) agree at the order-of-magnitude level with the bright-phase disk wind of \citet{Kosec2025} ($\log\xi \simeq 3.65$--$3.9$, $N_{\rm H} \simeq (0.7$--$2)\times10^{23}~{\rm cm^{-2}}$). The somewhat lower eclipse $N_{\rm H}$ reflects a systematic uncertainty: it scales inversely with the assumed solid angle ($\Omega/4\pi = 0.1$; \citealt{Kosec2020}), so a smaller value would shift it toward the absorption measurements. Both therefore trace the same clumpy wind.}
 
We model the local clump density as $n(r) \propto r^{-2}$, where $r$ is the radial distance from the neutron star, with a volume filling factor $C_V$ characterizing the clumpiness of the wind. Following \citet{Kosec2020}, we estimate the mass outflow rate by treating $C_V n(r)$ as the effective mean density of the clumpy wind. The hydrogen column density is then $N_{\rm H} = \int_{R_{in}}^{\infty} C_V n(r)\,dr$. Substituting $n(r) = n(R_{in})(r/R_{in})^{-2}$ and evaluating the integral gives $N_{\rm H} = C_V n(R_{in}) R_{in}^2 \int_{R_{in}}^{\infty} r^{-2}\,dr = C_V n(R_{in}) R_{in}$, so that $C_V n(R_{in}) = N_{\rm H} / R_{in}$. The mass outflow rate at the inner boundary is therefore \begin{equation} \dot{M}_{\rm out} = \mu m_p\,\Omega\,R_{in}^2\,C_V n(R_{in})\,v \approx \mu m_p \Omega N_{\rm H} R_{in} v, \end{equation} where $N_H = 3.1^{+0.9}_{-0.6} \times 10^{22}~{\rm cm^{-2}}$ is the hydrogen column density measured in this study (Section~\ref{sec3.4}), $\Omega/4\pi \approx 0.1$ is the covering fraction of the biconical wind (assuming $\Theta_0 \approx 5^\circ$; \citealt{Kosec2020}), and $v \approx 850$~km~s$^{-1}$ is the characteristic outflow velocity for a Compton-heated wind \citep{Kosec2020}. Because the luminosity $L$ and ionization parameter $\xi$ do not appear in this expression, the estimate is independent of the systematic uncertainties in these quantities.

This formulation yields $\dot{M}_{\rm out} \approx 3.2 \times 10^{-9}~M_{\odot}~{\rm yr^{-1}}$, approximately half of the mass accretion rate from the companion star, $\dot{M}_{\rm acc} \sim 7 \times 10^{-9}~M_{\odot}~{\rm yr^{-1}}$ \citep{Boroson2007}. The resulting net accretion rate onto the neutron star, $\dot{M}_{\rm true} \equiv \dot{M}_{\rm acc} - \dot{M}_{\rm out} \approx 3.8 \times 10^{-9}~M_{\odot}~{\rm yr^{-1}}$, implies a gravitational accretion luminosity of \begin{equation}   L_{\rm acc} = \frac{G M_X \dot{M}_{\rm true}}{R_X} \approx 5.1 \times 10^{37}~{\rm erg~s^{-1}}, \end{equation} where $M_X = 1.6~M_\odot$ and $R_X = 10$~km are the mass and radius of the neutron star, respectively. The 0.1--100~keV bolometric luminosity of Her~X-1, $L_{\rm bol} \approx 4.9 \times 10^{37}~{\rm erg~s^{-1}}$, derived from NuSTAR spectral fitting \citep{Wolff2016}, is in good agreement with $L_{\rm acc}$. These results indicate that X-ray irradiation efficiently drives the clumpy accretion-disk wind; the mechanical power of the outflow, $L_{\rm out} = \frac{1}{2}\dot{M}_{\rm out}v^2 \approx 7 \times 10^{32}~{\rm erg~s^{-1}}$, requires less than $5 \times 10^{-5}$ of the observed X-ray luminosity.

Finally, the derived $R_{\rm in}$ also allows us to independently constrain the wind number density at the launching point. The inferred value of $n(R_{\rm in}) \approx 2 \times 10^{13}~\rm{cm^{-3}}$ is consistent with the lower limit of $n \gtrsim 10^{12}~{\rm cm^{-3}}$ derived from pulse-resolved analysis \citep{Kosec2024}. At such high densities, the variation of the wind ionization state in response to X-ray pulsations is expected to be detectable with XRISM, although the response timescale is too short to directly measure the number density from the timing of the ionization response.

\section{Conclusions} \label{sec:conclusion}
\added{Using XRISM/Resolve with $\sim5$~eV resolution, we present eclipse-resolved Fe-K spectroscopy of Her~X-1 that separates the neutral and highly ionized components, constraining the disk kinematics and the properties of the reprocessing plasma.} Our main results are summarized as follows:
\begin{itemize}
\item The neutral Fe~K$\alpha$ line is not detected during mid-eclipse, and its centroid shifts from red at ingress to blue at egress. This sign reversal is a direct signature of rotation in the outer-disk fluorescing material ($v_{\rm disk}\sim180$~km~s$^{-1}$), suggesting an association with the outer accretion disk.
\item Eclipse flux-ratio diagnostics constrain the geometric inner boundary of the clumpy disk wind to $R_{\rm in} = 3^{+5}_{-2} \times 10^{10}$~cm ($1\sigma$), falling within the theoretically expected wind-launching window of ${\sim}(0.8\text{--}8)\times10^{10}$~cm, consistent with a thermally driven origin.
\item The implied mass outflow rate is $\dot{M}_{\rm out} \approx 3.2 \times 10^{-9}~M_\odot~{\rm yr}^{-1}$, approximately half of the mass supply rate from the companion star. The net accretion luminosity of $L_{\rm acc} \approx 5.1 \times 10^{37}$~erg~s$^{-1}$ is consistent with the observed bolometric luminosity of Her~X-1.
\end{itemize}

\begin{acknowledgments}
This research made use of data obtained with XRISM, a JAXA, NASA, and ESA mission. Data analysis tools are provided by HEASoft and the XRISM CALDB.
T.E. acknowledges support from JSPS KAKENHI Grant Number JP26H02075. 
We thank M. Tsujimoto, Y. Mochizuki, and N. Sameshima (JAXA/ISAS), Y. Nagai, K. Matsunaga, S. Inoue (Kyoto University) for helpful comments and discussions during the preparation of this work.
\end{acknowledgments}

\appendix
\section{3D Eclipse Geometry and Analytical Emission Measure}
\label{app:A}
\added{The goal of this appendix is to express the observed emission-line flux ratio between eclipse intervals in terms of the wind's inner radius $R_{\rm in}$. The line flux at any instant is proportional to the emission measure of the wind that remains visible (unmasked by the companion), so for each interval (ingress, mid-eclipse, egress) we require the interval-averaged emission measure
\begin{equation}
\langle\mathcal{EM}\rangle_{\rm interval} = \frac{1}{\Delta t_{\rm interval}} \int_{\rm interval} dt \int_{V_{\rm vis}(t)} dV\,n^2,
\end{equation}
that is, the squared wind density integrated over the visible volume $V_{\rm vis}(t)$ and averaged over the interval duration $\Delta t_{\rm interval}$. We ultimately invert the ratio $\eta_{\rm line}=\langle\mathcal{EM}\rangle_{\rm ing}/\langle\mathcal{EM}\rangle_{\rm mid}$, in which the common prefactors cancel, leaving a dependence on $R_{\rm in}$ alone. The remainder of the appendix evaluates this double integral by reducing the spatial integral to a line-of-sight profile and mapping the time integral onto the companion's sweep across the wind (\autoref{ponti}c).}

We adopt a Cartesian coordinate system $(x, y, z)$ centered on the neutron star, with the observer along $+x$ and the companion star moving along the $y$-axis (see \autoref{ponti}c). We define $r \equiv \sqrt{x^2+y^2+z^2}$ and $R \equiv \sqrt{x^2+y^2}$ for the spherical and cylindrical radii, respectively. We model the wind with an $r^{-2}$ number density profile\added{, as expected for a steady, constant-velocity outflow,} truncated at the cylindrical inner boundary $R = R_{\rm in}$ and confined to a biconical equatorial wedge $|z| \le R\tan\Theta_0$\added{, representing a wind launched into a narrow range of angles above and below the disk plane}:

\begin{equation}
n(x, y, z) =
\begin{cases}
  n_0\,\dfrac{R_{\rm in}^2}{x^2+y^2+z^2} & (R \ge R_{\rm in}),\\[6pt]
  0 & (R < R_{\rm in}),
\end{cases}
\end{equation}
where $n_0$ is the normalization density at $r = R_{\rm in}$.
Integrating $n^2$ along $z$ with the substitution $z = R\tan\alpha$ yields the surface emission-measure density

\begin{equation}
\Sigma_{\mathcal{EM}}(x, y)
  = \frac{n_0^2 R_{\rm in}^4}{(x^2+y^2)^{3/2}}\,C_0,
  \qquad C_0 \equiv \Theta_0 + \sin\Theta_0\cos\Theta_0.
\end{equation}

Projecting $\Sigma_{\mathcal{EM}}$ along the line of sight ($x$-axis) and evaluating with $\int(x^2+y^2)^{-3/2}dx = x/(y^2\sqrt{x^2+y^2})$ gives the 1D linear emission-measure profile

\begin{equation}
\Lambda_{\mathcal{EM}}(y) =
\begin{cases}
  \dfrac{2\,C_0\,n_0^2 R_{\rm in}^4}{y^2}
    & (y \ge R_{\rm in}),\\[8pt]
  \dfrac{2\,C_0\,n_0^2 R_{\rm in}^3}{R_{\rm in}+\sqrt{R_{\rm in}^2-y^2}}
    & (y < R_{\rm in}),
\end{cases}
\end{equation}

where the inner-region form is obtained by multiplying through by the conjugate $(R_{\rm in}+\sqrt{R_{\rm in}^2-y^2})$ to remove the singularity at $y=0$.

We define the cumulative emission measure
\begin{equation}
\mathcal{EM}(s) \equiv \int_s^{\infty}\Lambda_{\mathcal{EM}}(y')\,dy',
\end{equation}
\added{where $y'$ is a dummy variable of integration along the projected coordinate $y$, and $s$ is the masking depth, defined as the projected distance by which the near limb of HZ~Herculis has swept past the neutron star (see \autoref{ponti}c).} Thus $\mathcal{EM}(s)$ is the cumulative emission measure contributed by wind at projected $y>s$; for $s \ge R_{\rm in}$ this reduces to $\mathcal{EM}(s) = 2\,C_0 n_0^2 R_{\rm in}^4/s$.

Because HZ~Herculis has a finite radius $R_\ast$, the eclipse produces two distinct visible wind regions at any masking depth $s \geq 0$. When the near limb has swept a distance $s$ past the neutron star, it sits at $y=-s$, and the far limb sits at $y=2R_\ast-s$; the companion occults the band $y\in[-s,\,2R_\ast-s]$. Using the even symmetry of $\Lambda_{\mathcal{EM}}$, the near-side wind ($y<-s$) contributes $\mathcal{EM}(s)$, and the far-side wind ($y>2R_\ast-s$) contributes $\mathcal{EM}(2R_\ast-s)$, giving the exact instantaneous cumulative emission measure

\begin{equation}
\mathcal{EM}_{\rm exact}(s) = \mathcal{EM}(s) + \mathcal{EM}(2R_\ast-s).
\end{equation}
\added{The finite size of HZ~Herculis is essential here: at any instant the companion hides a band of the wind, leaving two strips visible simultaneously, one on the near side of the neutron star and one on the far side, so that the observed emission is the sum of these two contributions.}

We map the observed 1:2:1 duration ratio of the ingress, mid-eclipse, and egress intervals onto ranges of $s$ by assuming a constant orbital velocity. \added{Over the full eclipse the near limb sweeps the masking depth from $s=0$ to $s=2R_\ast$, so the 1:2:1 split assigns ingress to $s\in[0,\,R_\ast/2]$, mid-eclipse to $s\in[R_\ast/2,\,3R_\ast/2]$, and egress to $s\in[3R_\ast/2,\,2R_\ast]$. Since $\mathcal{EM}_{\rm exact}(s)$ is symmetric about $s=R_\ast$, the mid-eclipse range is equivalent to $s\in[R_\ast/2,\,R_\ast]$ traversed twice, and egress is identical to ingress.} We assume $R_\ast/2 \gg R_{\rm in}$ throughout, so that $\mathcal{EM}(s) = 2\,C_0 n_0^2 R_{\rm in}^4/s$ applies over all integration ranges involving the far-side term. The three interval-averaged cumulative emission measures are then

\begin{align}
\langle \mathcal{EM} \rangle_{\rm mid}
  &= \frac{2}{R_\ast}
     \int_{R_\ast/2}^{R_\ast}
     \bigl[\mathcal{EM}(s)+\mathcal{EM}(2R_\ast-s)\bigr]\,ds
   = \frac{4\,C_0 n_0^2 R_{\rm in}^4}{R_\ast}\,\ln 3, \\[4pt]
\langle \mathcal{EM} \rangle_{\rm ing}
  &= \frac{2}{R_\ast}
     \int_0^{R_\ast/2}
     \bigl[\mathcal{EM}(s)+\mathcal{EM}(2R_\ast-s)\bigr]\,ds
   = \frac{4\,C_0 n_0^2 R_{\rm in}^4}{R_\ast}
     \left[2+\ln\!\left(\frac{R_\ast}{3R_{\rm in}}\right)\right], \\[4pt]
\langle \mathcal{EM} \rangle_{\rm egr}
  &= \langle \mathcal{EM} \rangle_{\rm ing}.
\end{align}

The mid-eclipse integral uses $\int_{R_\ast/2}^{R_\ast}s^{-1}ds = \ln 2$ together with $\int_{R_\ast}^{3R_\ast/2}u^{-1}du = \ln(3/2)$ for the far-side term (after $u=2R_\ast-s$). The ingress integral combines the near-side contribution, evaluated by integration by parts with $s=R_{\rm in}\sin\theta$ over $s < R_{\rm in}$, with the far-side correction $\int_{3R_\ast/2}^{2R_\ast}u^{-1}du = \ln(4/3)$.

To improve the signal-to-noise ratio of the observed line fluxes, ingress and egress spectra are combined into a single time-averaged measurement $\langle F\rangle_{\rm ing+egr}$. Because the two intervals have equal duration and equal cumulative emission measure ($\langle \mathcal{EM}\rangle_{\rm ing} = \langle \mathcal{EM}\rangle_{\rm egr}$), this average equals the individual interval value, and the observed line flux ratio is

\begin{equation}
\eta_{\rm line} \equiv
  \frac{\langle F \rangle_{\rm ing+egr}}{\langle F \rangle_{\rm mid}}
= \frac{\langle \mathcal{EM} \rangle_{\rm ing}}{\langle \mathcal{EM} \rangle_{\rm mid}}
= \frac{2+\ln\!\left(R_\ast/3R_{\rm in}\right)}{\ln 3}.
\end{equation}

Inverting for $R_{\rm in}$:

\begin{equation}
R_{\rm in}
  = \frac{R_\ast}{3}
    \exp\!\left[2 - \eta_{\rm line}\cdot\ln 3\right].
\end{equation}
\added{Intuitively, a more compact wind (smaller $R_{\rm in}$) loses a larger fraction of its flux once the companion masks its center, giving a larger $\eta_{\rm line}$; measuring $\eta_{\rm line}$ thus inverts directly to the wind's inner radius.}
We apply this formulation to the observed highly ionized Fe lines in Section~\ref{42}.

\bibliography{reference}
\bibliographystyle{aasjournalv7}

\end{document}